\documentclass[a4paper,12pt]{article}
\usepackage[english]{babel}
\usepackage[utf8]{inputenc}
\usepackage{cite}
\usepackage{amsthm,amsfonts,amsmath,mathrsfs}
\usepackage{slashed}
\usepackage{makeidx}
\usepackage{graphicx}
\usepackage{caption}
\usepackage{hyperref}
\usepackage{simplewick}
\usepackage{verbatim}
\usepackage{natbib}

\newcommand{\be}{\begin{eqnarray}}
	\newcommand{\ee}{\end{eqnarray}}

\providecommand{\keywords}[1]
{
  \small	
  \textbf{\textit{Keywords}} #1
}

\begin{document}
	
	\title{Propagation features of Lorentz-violating electrodynamics} 
	
	\maketitle
	\begin{center}
	  E. Goulart$^a$, J. E. Ottoni$^{a}$, J. C. C. Felipe$^a$
	\\[2em]
	{\sl ${}^a$ Departamento de Estat\' istica, F\'isica e Matem\'atica, Universidade Federal de São João del Rei, Rod. MG 443, Km 7, 36497-899 - Ouro Branco - MG, Brasil}\\
	\end{center}
	
	\begin{abstract}
	\noindent	
    This paper explores some propagation features of electrodynamics in a Lorentz-violating scenario, focusing on a specific CPT-even term within the photon sector of the Standard Model Extension (SME). The study derives a covariant dispersion relation for light propagation in the presence of a Lorentz-violating symmetric tensor field \(C_{ab}(x)\), which reveals a modified light cone structure described by a quartic polynomial. The analysis includes a simplified model where the tensor field assumes a dyadic form, leading to an effective metric that dictates the propagation. The paper investigates three distinct cases: Lorentz-violating vectors which are timelike, lightlike and spacelike, and examines their implications for the effective velocity of light, birefringence, and analogies with electrodynamics in material media. The results highlight anisotropic propagation effects and provide insights into the interplay between Lorentz violation and modified causal structures. The study concludes with a discussion of the phenomenological implications and possible experimental tests for such Lorentz-violating effects.  

	\end{abstract}

	
	\section{Introduction}

\quad Lorentz and CPT symmetries are fundamental cornerstones that underlie our description of the behavior of particles and fields. These symmetries are expected to be satisfied in any physical situation and lie at the heart of the Standard Model of particle physics (SM). To date, the overwhelming majority of experimental searches for violations of these symmetries have yielded negative results. However, in the last few decades, the possibility of spontaneously breaking Lorentz and CPT symmetries has been extensively suggested by various theoretical constructs such as string theory, loop quantum gravity, and non-commutative field theories. From a broader perspective, the general framework adequate for these studies is an effective field theory called the Standard Model Extension (SME). The main idea is that the low-energy effective action appearing in the extension involves the expectation of a tensor field in the underlying theory, thus engendering the appropriate symmetry breakdowns (see, for instance, \cite{kost1,colladay1997cpt,colladay1998lorentz,kostelecky2004gravity}).

Among all the different sectors proposed by the SME, the photon sector turns out to be of huge importance: since most of what we infer about the large-scale structure of the universe is based on the propagation of light itself, even tiny deviations of the SM assumptions could lead to drastic experimental consequences. Furthermore, in extreme field environments, such as high-energy astrophysical phenomena or primordial cosmological settings one expects to encounter other possible extensions of Maxwell's theory. The latter may include nonlinear electrodynamics, non-minimal couplings of the electromagnetic field with gravity, area metric theories and others (\cite{sorokin2022introductory},\cite{balakin2005non},\cite{punzi2009propagation}). The problem we face is that all the above extensions predict different dispersion relations which give rise to backgrounds as refinements of the ordinary metric geometry. Therefore, in order to test Lorentz-breaking electrodynamics with precision, it is mandatory that we manage to separate and categorize all possible effects predicted by the different theories (the reader is invited to consult \cite{perlick2011hyperbolicity},\cite{Raetzel:2010je},\cite{de2000light},\cite{abalos2015nonlinear},\cite{de2009classification},\cite{Goulart:2021uzr} for some of these effects). 

In this paper, we focus our attention into a class of Lorentz-violating electrodynamics within the CPT-even photon sector of the SME. It is worth mentioning that the interest in this sector has increased some time ago in connection with the \textit{aether} concept and extra dimensions. In \cite{Carroll}, for instance, Lorentz-violating tensor fields with expectation values aligned with the extra dimensions were used so as to keep these extra dimensions hidden. The Lorentz-violating terms of the gauge, fermion and scalar sectors, in this case, would come from the interaction between these fields with the \textit{aether} fields. More generally, the kind of CPT-even term we are considering can be radiatively generated when a Lorentz-violating non-minimal coupling is introduced in electrodynamics \cite{Petrov,Scarp,Scarp2,Petrov-Scarp}.

More specifically, we consider a CPT-even term controlled by a general symmetric tensor field $C_{ab}(x)$. The latter gives rise to a constitutive tensor of ten parameters that allows us to derive a fully covariant dispersion relation in explicit form, thus providing some insights into the propagation of light in the presence of Lorentz-violating terms. It is shown that the dispersion relation is necessarilly of fourth order in the wave covector and contains the symmetric tensor up to third order. Then, restricting ourselves to the case where $C_{ab}(x)$ has a simple dyadic structure, we show that the quartic polynomial factorizes and analyze the behavior of the corresponding cones. To complete the picture, we explore connections between our results and the electrodynamics of light inside material media, where similar effects can arise due to the interaction with the medium. These connections offer a unique opportunity to explore the interplay between fundamental physics and condensed matter physics and may be used to test the theory in a laboratory framework.



As mentioned before, analyzing aspects of Lorentz-violation from the experimental point of view may be quite tricky. Fortunately, there is an extensive compilation of constraints on the values of the coefficients for Lorentz and CPT violation in the SME \cite{Kost5,Kost-CPT}, allowing us to somehow quantify the violations arising from the diﬀerent sectors. Particular examples include multi-messenger astronomy \cite{Adda}, gravitational waves\cite{Kost2,Kost3,Schr} cosmic rays \cite{Kost4,Alts} and astrophysical tests with neutrinos and gamma-ray photons \cite{Bert} (see also \cite{kost1,Dopli,Boj,Amel}). We hope that our results may help in separating the effects generated by Lorentz-violating electrodynamics and other extensions of Maxwell electrodynamics.




This paper is organized as follows: On chapter two, we present the Lagragian of the model and this respectivelly equation of motion. On chapter three we deduce the dispersion relations in a covariant way. In chapter four,  we discusse the model in a material media scenario. On chapter five, we analyse the Lorentz violation term as a spacelike, null and timelike vector. So, on chapter six, we present the conclusions and perspectives.

\section{Lagrangian and equations of motion}

To begin with, we let $(M,g_{ab})$ denote a four-dimensional Minkowski spacetime with metric signature convention $(+,-,-,-)$. In an arbitrary coordinate system $\{x^{a}\}$, the electromagnetic field tensor and its Hodge dual are expanded, respectively, as
\begin{equation}
\boldsymbol{F}=\frac{1}{2}F_{ab}\ dx^{a}\wedge dx^{b},\quad\quad\quad \star\boldsymbol{F}=\frac{1}{2}\star F_{ab}\ dx^{a}\wedge dx^{b},
\end{equation}
where
\begin{equation}
\star F_{ab}=\frac{1}{2}\varepsilon_{ab}^{\phantom a\phantom a pq}F_{pq},
\end{equation}
with the Levi-Civita tensor given by $\varepsilon_{abcd}=\sqrt{-g}\ [abcd]$ and the permutation symbol defined such that $[0123]\equiv +1$. In general, given a field of observers, say $t^{q}(x)$, in spacetime, the field strength uniquely decomposes as\footnote{Throughout, we stick to the non-normalized convention for symmetrization and anti-symmetrization i.e., $(ab)\equiv ab+ba$ and $[ab]\equiv ab-ba$.}
\begin{eqnarray}\label{strength}
F^{ab}&=&E^{[a}t^{b]}+\varepsilon^{ab}_{\phantom a\phantom a cd}B^{c}t^{d}
\end{eqnarray}
where the projections
\begin{equation}
E^{a}=F^{a}_{\phantom a b}t^{b},\quad\quad B^{a}=-\star F^{a}_{\phantom a b}t^{b},
\end{equation}
stand for the electric and magnetic fields as measured by this particular observer. There follows that these vectors are spacelike and orthogonal to $t^{q}(x)$ by construction and we notice the dual tensor may be easily obtained in terms of the simple permutations $E^{a}\mapsto -B^{a}$ and $B^{a}\mapsto E^{a}$ in Eq. (\ref{strength}).

Motivated by the study of electrodynamics in a Lorentz-breaking scenario, we shall consider the following gauge-invariant CPT-even lagrangian density
\begin{equation}\label{ld}
\mathcal{L}=-\frac{1}{4}F_{ab}F^{ab}-\frac{1}{2}C_{ab}F^{ac}F^{b}_{\phantom a c},
\end{equation}
where $F_{ab}=\partial_{a}A_{b}-\partial_{b}A_{a}$ as usual and $C_{ab}(x)$ is a smooth symmetric tensor field depending solely on spacetime position. Latter on, for the sake of simplicity, we shall restrict the latter to a simple dyadic structure involving a single vector field. However, at this point, we keep this tensor in its general algebraic form and define the useful irreducible quantities
\begin{equation}
C=C^{a}_{\phantom a a},\quad\quad\quad\hat{C}^{ab}=C^{ab}-\frac{1}{4}Cg^{ab}.
\end{equation}

It turns out that it is more convenient for our purposes here to make some algebraic manipulations in order to  rewrite Eq. (\ref{ld}) in the general form
\begin{equation}\label{action}
\mathcal{L}=-\frac{1}{8}X_{abcd}F^{ab}F^{cd}
\end{equation}
with the double form defined by
\begin{equation}\label{df}
X_{abcd}=Ng_{abcd}+g_{[a[c}\hat{C}_{b]d]},
\end{equation}
where $N=1+C/2$ for conciseness, $g_{abcd}=g_{ac}g_{bd}-g_{ad}g_{bc}$ is the bi-metric tensor and anti-symmetrization on the r.h.s. is supposed to act on the pairs $ab$ and $cd$ individually. We notice that Eq. (\ref{df}) corresponds to the irreducible decomposition of $X_{abcd}$ under the pseudo-orthogonal group $O(1,3)$ from which one concludes that the double form has a total of $10$ independent components. Also, there follow the algebraic symmetries
\begin{equation}
X_{abcd}=-X_{bacd},\quad X_{abcd}=-X_{abdc},\quad X_{abcd}=X_{cdab},\quad X_{a[bcd]}=0.
\end{equation}
Consequently, this tensor is characterised as a particular instance of a symmetric curvaturelike double form in four dimensions. We refer the reader to \cite{goulart2023remarks} where a wealth of details concerning the algebraic structure of double forms is discussed.

With the above conventions, variation of the action Eq. (\ref{action}) with respect to the four-potential gives, in a source-free region of spacetime, the following equations of motion\footnote{Since we are mainly interested in the propagation aspects of the theory, the inclusion of a source term is unimportant for our conclusions.}
\begin{equation}\label{eom}
(X^{ab}_{\phantom a\phantom a cd}F^{cd})_{;b}=0,\quad\quad\quad F_{ab;c}+F_{ca;b}+F_{bc;a}=0,\end{equation}
where $;$ denotes covariant derivative compatible with the metric i.e., $g_{ab;c}=0$. Since the double form $X_{abcd}$ is assumed to depend solely on spacetime position (but not on the electromagnetic field),  Eqs. (\ref{eom}) constitute a linear system of first order partial differential equations for the field strength. Furthermore, since we have a total of eight equations for only six unknowns, one expects that the set of equations split into two constraints plus six evolution equations.

\section{Covariant dispersion relation}

\subsection{Generalities}
As is well known, the dispersion relation associated to Eqs. (\ref{eom}) may be obtained either using Hadamard's method of weak discontinuities or the eikonal approximation, which we now apply.  We start by assuming an approximate wavy solution to Eqs. (\ref{eom}) of the form
\begin{equation}
F_{ab}\approx f_{ab}(x)e^{i\Theta(x)},
\end{equation}
with $f_{ab}(x)$ a slowly varying amplitude and $\Theta(x)$ a rapidly varying phase. In this approximation we neglect gradients in the amplitude and retain only the gradients of the phase $k_{a}\equiv\partial_{a}\Theta$. A simple calculation then shows that the Bianchi identity reduces to the algebraic relation 
\begin{equation}
f_{ab}k_{c}+f_{ca}k_{b}+f_{bc}k_{a}=0,
\end{equation}
from which one concludes that the amplitude of the electromagnetic disturbance reduces to a simple 2-form, which may be written as
\begin{equation}\label{Bianchi}
f_{ab}=k_{a}a_{b}-k_{b}a_{a},
\end{equation}
with $a_{a}$ denoting the polarization 1-form. Applying Eq. (\ref{Bianchi}) to the first equation in Eqs.\ (\ref{eom}) gives the complementary condition
\begin{equation}\label{main}
(X^{ambn}k_{m}k_{n})a_{b}=0,
\end{equation}
which is the building block of the covariant dispersion relation: it implies an algebraic constraint which must be fulfilled by the characteristic covectors $k_{a}$ in order to obtain a physically meaningful solution. 

In order to investigate the algebraic implications of Eq.\ (\ref{main}) in more details, we shall fix a point $x$ on the manifold and consider the generic nonlinear map
\begin{equation}\label{ymap}
Y: T_{x}^{*}M\rightarrow\mbox{Mat}_{4\times 4}(\mathbb{R}),\quad\quad q_{m}\mapsto Y^{ab}\equiv X^{ambn}q_{m}q_{n}.
\end{equation}
An important property of the latter is that every covector in the domain produces a matrix which automatically annihilates the corresponding covector. In other words, we have
\begin{equation}\label{algcond}
Y^{ab}q_{b}=Y^{ba}q_{b}=0.
\end{equation}
This means that the image of $T_{x}^{*}M$ in the space $\mbox{Mat}_{4\times 4}(\mathbb{R})$ belongs to the determinantal variety defined by $\mbox{rk}\ Y^{ab}\leq 3$. When combined with the Cayley-Hamilton theorem, this fact guarantees that
\begin{equation}
Y^{a}_{\phantom a c}Z^{c}_{\phantom a b}=Z^{a}_{\phantom a c}Y^{c}_{\phantom a b}=0,
\end{equation}
where
\begin{eqnarray}\label{adjointmatrix}
Z^{a}_{\phantom a b}&\equiv&Y^{a}_{\phantom a c}Y^{c}_{\phantom a d}Y^{d}_{\phantom a b}-\sigma_{1}Y^{a}_{\phantom a c}Y^{c}_{\phantom a b}+\sigma_{2}Y^{a}_{\phantom a b}-\sigma_{3}\delta^{a}_{\phantom a b},
\end{eqnarray}
is the classical adjoint with the following elementary symmetric polynomials 
\begin{equation}
\sigma_{1}\equiv Y^{p}_{\phantom a p},\quad\quad\quad \sigma_{2}\equiv\frac{1}{2}(Y^{p}_{\phantom a p}Y^{q}_{\phantom a q}-Y^{p}_{\phantom a q}Y^{q}_{\phantom a p}),
\end{equation}
\begin{equation}
\sigma_{3}\equiv\frac{1}{6}(Y^{p}_{\phantom a p}Y^{q}_{\phantom a q}Y^{r}_{\phantom a r}-3Y^{p}_{\phantom a p}Y^{q}_{\phantom a r}Y^{r}_{\phantom a q}+2Y^{p}_{\phantom a q}Y^{q}_{\phantom a r}Y^{r}_{\phantom a p}).
\end{equation}
Clearly, the classical adjoint is a cubic combination of the tensor $X^{abcd}$ and is of sixth order in the covector $q_{m}$, but a well known result of linear algebra in four dimensions states that:
\begin{enumerate}
\item{when $\mbox{rk}\ Y^{ab}=3$, then $\mbox{rk}\ Z^{ab}=1$;}\\
\item{when $\mbox{rk}\ Y^{ab}<3 $, then $Z^{ab}=0$.}
\end{enumerate}

To begin with, let us first assume that the image of a given covector $q_{a}$ in the space $\mbox{Mat}_{4\times 4}(\mathbb{R})$ belongs to the upper bound of the determinantal variety i.e., assume that $\mbox{rk}\ Y^{a}_{\phantom a b}=3$. In this case, simple algebraic manipulations using Eq.\ (\ref{algcond}) show that the adjoint must have a trivial dyadic structure of the form\\
\begin{equation}\label{Zed}
Z_{ab}= H(x,q)q_{a}q_{b},
\end{equation}\\
with\\
\begin{equation}
H(x,q)\equiv G^{abcd}(x)q_{a}q_{b}q_{c}q_{d},
\end{equation}\\
for some symmetric fourth order contravariant tensor. Now, suppose that we continuously vary the covector $q_{a}$ until we achieve a new covector in the cotangent space, say $k_{a}$, such that its image belongs to the sub-variety defined by $\mbox{rk}\ Y^{ab}<3$. Since $Z_{ab}$ vanishes for this case, due to continuity arguments, we must impose $H(x,k)=0$, i.e.\\
\begin{equation}\label{disprel}
G^{abcd}(x)k_{a}k_{b}k_{c}k_{d}=0.
\end{equation}\\
Up to an unimportant conformal factor, this equation has the form of the dispersion relation we are looking for. In other words, a necessary condition for Eq.\ (\ref{main}) to admit nontrivial solutions for the polarization one-form at a spacetime point is that the corresponding wave covectors belong to  the vanishing set of a quartic polynomial in the cotangent space. Interestingly, it can be shown that

\begin{equation}\label{hamilt}
H(x,k)\sim (X^{aj_{1}bj_{2}}{}^{\star}X^{\star}_{j_{1}k_{1}j_{2}k_{2}}X^{ck_{1}dk_{2}})k_{a}k_{b}k_{c}k_{d},
\end{equation}\\
where ${}^{\star}X^{\star}_{abcd}$ is the double Hodge dual of $X_{abcd}$. For a more rigorous demonstration in the context of premetric electrodynamics, the reader is invited to consult \cite{hehl2012foundations, itin2009light, lindell2005electromagnetic, favaro2012recent}.

\subsection{Explicit derivation}
In order to explicitly compute the characteristic polynomial Eq. (\ref{hamilt}) associated to the Lorentz-breaking scenario provided by Eq. (\ref{df}), some index gymnastics is needed. And since this derivation is somehow cumbersome, we present it here with some details. To begin with, it is convenient to define the first and second order contractions
\begin{equation}\label{defslm}
l^{a}=\hat{C}^{ab}k_{b},\quad\quad m^{a}=\hat{C}^{ab}l_{b} 
\end{equation}
and the coefficient $M=Nk^{2}+(k.l)$. With these conventions, we get\\
\begin{equation}
X^{aj_{1}bj_{2}}k_{a}k_{b}=Mg^{j_{1}j_{2}}-Nk^{j_{1}}k^{j_{2}}-k^{(j_{1}}l^{j_{2})}+k^{2}\hat{C}^{j_{1}j_{2}}.
\end{equation}\\
To compute the contraction of the latter with the appropriate double dual tensor, it is convenient to use the Ruse-Lanczos identity (see \cite{goulart2023remarks}) to write\\
\begin{equation}
{}^{\star}X^{\star}_{j_{1}k_{1}j_{2}k_{2}}=-Ng_{j_{1}k_{1}j_{2}k_{2}}+g_{[j_{1}[j_{2}}\hat{C}_{k_{1}]k_{2}]}.
\end{equation}
Some additional algebraic manipulations then give the linear combination of six different types of terms
\begin{eqnarray}\nonumber
(X^{aj_{1}bj_{2}}k_{a}k_{b}){}^{\star}X^{\star}_{j_{1}k_{1}j_{2}k_{2}}&=&A g_{k_{1}k_{2}}+Bk_{k_{1}}k_{k_{2}}+Ck_{(k_{1}}m_{k_{2})}+Dl_{k_{1}}l_{k_{2}}\\\label{firstcontract}
&+&E\hat{C}_{k_{1}k_{2}} +F\hat{C}_{k_{1}j_{1}}\hat{C}^{j_{1}}_{\phantom a k_{2}},
\end{eqnarray}
with the coefficients given by
\begin{eqnarray}
&& A=-2MN-2l^{2}+k^{2}\hat{C}_{ab}\hat{C}^{ab},\quad B=-N^{2},\quad C=1,\quad D=2,\\\nonumber\\\quad 
&&\quad\quad\quad\quad\quad\quad\quad\quad  E=2Nk^{2},\quad F=-2k^{2},
\end{eqnarray}
for conciseness. Multiplying Eq. (\ref{firstcontract}) by $X^{ck_{1}dk_{2}}k_{c}k_{d}$, one notices that the terms proportional to $B$ and $C$ identically vanish, whereas the remaining terms read as follows
\begin{equation}
A(2M+Nk^{2}),
\end{equation}
\begin{equation}
D[Nk^{2}l^{2}-N(k.l)^{2}-l^{2}(k.l)+k^{2}(l.m)],
\end{equation}
\begin{equation}
E[-N(k.l)-2l^{2}+k^{2}\hat{C}^{a}_{\phantom a b}\hat{C}^{b}_{\phantom a a}],
\end{equation}
\begin{equation}
F[M\hat{C}^{a}_{\phantom a b}\hat{C}^{b}_{\phantom a a}-Nl^{2}-2(l.m)+k^{2}\hat{C}^{a}_{\phantom a b}\hat{C}^{b}_{\phantom a c}\hat{C}^{c}_{\phantom a a}].
\end{equation}
Combination of these terms, then gives the fourth order polynomial
\begin{eqnarray*}
&&H(x,k)\sim\left\{N^{3}-\frac{N}{2}\hat{C}^{p}_{\phantom a q}\hat{C}^{q}_{\phantom a p}+\frac{1}{3}\hat{C}^{p}_{\phantom a q}\hat{C}^{q}_{\phantom a r}\hat{C}^{r}_{\phantom a p}\right\}k^{4}+\left\{2N^{2}(l.k)+Nl^{2}-(l.m)\right\}k^{2}\\\\
&&\quad\quad\quad\quad\quad\quad\quad\quad\quad +(l.k)\left\{N(l.k)+l^{2}\right\}.
\end{eqnarray*}

\noindent Finally, replacing the first and second order contractions defined by Eqs. (\ref{defslm}), we obtain
\begin{eqnarray}\nonumber
H(x,k) &&\sim \left\{N^{3}-\frac{N}{2}\hat{C}^{p}_{\phantom a q}\hat{C}^{q}_{\phantom a p}+\frac{1}{3}\hat{C}^{p}_{\phantom a q}\hat{C}^{q}_{\phantom a r}\hat{C}^{r}_{\phantom a p}\right\}k^{4}\\\nonumber\\\nonumber
&&+\left\{2N^{2}\hat{C}^{ab}+N\hat{C}^{a}_{\phantom a p}\hat{C}^{pb}-\hat{C}^{ap} \hat{C}_{p}^{\phantom a q}\hat{C}_{q}^{\phantom a b}\right\}k_ak_bk^{2}\\\nonumber\\\label{hxq}
&&+\left\{N\hat{C}^{ab}+\hat{C}^{a}_{\phantom a r}\hat{C}^{rb}\right\}\hat{C}^{cd}k_ak_bk_ck_d.
\end{eqnarray}
It is clear that, at each cotangent space $T^{*}_{x}M$, the light cone associated to the Lorentz-breaking scenario is not given by the ordinary Minkowskian cone, but rather by a fourth order polynomial surface, so it can have multiple sheets and singular points. Before we proceed to the analysis of these cones in a simplified model, let us compare the above formalism with the electrodynamics inside linear and local media. 

\section{Analogy with material media}

When dealing with electromagnetic fields propagating inside local and linear media, one is often concerned with the local constitutive equation
\begin{equation}\label{lmap}
H^{ab}=\frac{1}{2}X^{ab}_{\phantom a\phantom a cd}F^{cd},
\end{equation}
where $H^{ab}$ is the excitation tensor and $X_{abcd}$ is a generic double form depending on position but not on the field strength. Such an object is called the constitutive tensor, has a total of $36$ independent components and works as a linear map between the space of field strengths into the space of field excitations. In this framework, particular types of media are obtained by imposing additional algebraic symmetries into the double form. 

In order to analyse what type of media corresponds to the Lorentz-breaking term described by Eq. (\ref{df}), we start by recalling the $(3+1)$ splitting of the relevant quantities. In general, given a field of observers $t^{q}(x)$ in spacetime, the excitation tensor decomposes as
\begin{eqnarray}
H^{ab}&=&D^{[a}t^{b]}+\varepsilon^{ab}_{\phantom a\phantom a cd}H^{c}t^{d}
\end{eqnarray}
with the projections
\begin{equation}
D^{a}=H^{a}_{\phantom a b}t^{b},\quad\quad H^{a}=-\star H^{a}_{\phantom a b}t^{b}.
\end{equation}
Here, the couple of spacelike vectors orthogonal to the observer at each point $\{D^{a},H^{a}\}$ stand for the corresponding electric and magnetic displacements. 

Similarly, due to the so called Bel-Matte decomposition (see \cite{goulart2023remarks}), the constitutive tensor uniquely splits as
\begin{eqnarray}\nonumber
X_{abcd}&=&\{g_{abpq}(g_{cdrs}\mathfrak{A}^{pr}-\varepsilon_{cdrs}\mathfrak{B}^{pr})\\\label{Beldec}
&+&\ \ \varepsilon_{abpq}(g_{cdrs}\mathfrak{C}^{pr}-\varepsilon_{cdrs}\mathfrak{D}^{pr})\}t^{q}t^{s},
\end{eqnarray}
with the $2$-index tensors given by\footnote{Here, we stick to the convention where the position of the star denotes the skew pair which is to be Hodge dualized.}
\begin{equation}\label{Beldec1}
\mathfrak{A}_{ac}=X_{abcd}t^{b}t^{d},\quad \mathfrak{B}_{ac} =X^{\star}_{abcd} t^{b}t^{d},
\end{equation}
\begin{equation}\label{Beldec2}
\mathfrak{C}_{ac}= -{}^{\star}X_{abcd} t^{b}t^{d},\quad \mathfrak{D}_{ac} =-{}^{\star}X^{\star}_{abcd} t^{b}t^{d}.
\end{equation}
Clearly, the \textit{constitutive tetrad} $\{\mathfrak{A}^{ab},\mathfrak{B}^{ab},\mathfrak{C}^{ab},\mathfrak{D}^{ab}\}$ is orthogonal to the vector $t^{q}$ by construction and each member has a total of $9$ independent components, as expected. With the above conventions, one readily sees that
\begin{eqnarray}
D^{a}=\mathfrak{A}^{a}_{\phantom a b}E^{b}+\mathfrak{B}^{a}_{\phantom a b}B^{b},\quad\quad H^{a}=\mathfrak{C}^{a}_{\phantom a b}E^{b}+\mathfrak{D}^{a}_{\phantom a b}B^{b}.
\end{eqnarray}
Here, $\mathfrak{A}^{a}_{\phantom a b}$ stands for the \textit{permittivity tensor} whereas $\mathfrak{D}^{a}_{\phantom a b}$ denote the \textit{inverse permeability tensor}. The cross terms $\mathfrak{B}^{a}_{\phantom a b}$ and $\mathfrak{C}^{a}_{\phantom a b}$ are the \textit{magnetoelectric tensors}, which describe a linear response of the electric polarization to a magnetic field, and vice versa. It is clear from these definitions that, in a general situation, the characteristic polynomial Eq. (\ref{hamilt}) can contain up to third order combinations of the constitutive tetrad.

Now, since the mathematical structure of Eq. (\ref{lmap}) and that provided by the Lorentz breaking scenario are essentially the same, we get the following constitutive tetrad associated to (\ref{df})
\begin{eqnarray}
\mathfrak{A}_{ac}&=&(N+\hat{C}_{pq}t^{p}t^{q})p_{ac}+{}^{(3)}\hat{C}_{ac},\\\nonumber\\
\mathfrak{B}_{ac}&=&{}^{(3)}\varepsilon_{ac}^{\phantom a\phantom a p}\hat{C}_{pq}t^{q}\\\nonumber\\
\mathfrak{C}_{ac}&=&{}^{(3)}\varepsilon_{ac}^{\phantom a\phantom a p}\hat{C}_{pq}t^{q}\\\nonumber\\
\mathfrak{D}_{ac}&=&(N-\hat{C}_{pq}t^{p}t^{q})p_{ac}-{}^{(3)}\hat{C}_{ac}
\end{eqnarray}
where
\begin{equation}
p_{ab}\equiv g_{ab}-t_{a}t_{b},\quad\quad {}^{(3)}\hat{C}_{ab}\equiv p_{a}^{\phantom a p}p_{b}^{\phantom a q}\hat{C}_{pq},\quad\quad {}^{(3)}\varepsilon_{abc}\equiv t^{p}\varepsilon_{pabc},
\end{equation}
stand for the projector induced by the observer field, the projection of $C_{ab}$ onto the rest space of $t^{q}$, and the three-dimensional Levi-Civita tensor,  respectively. From the latter, one realizes that the permittivity and inverse permeability tensors are symmetric whereas the magnetoelectric cross terms are equal and antisymmetric.

\section{A simplified model}

As mentioned before, the vanishing sets of the characteristic polynomial Eq. (\ref{hxq}) can have a quite complex geometric structure for an arbitrary tensor field $C_{ab}(x)$. In order to investigate these sets in more details, we consider a simplified model where $\mbox{rank}(C_{ab})=1$. More precisely, we consider the following \textit{ansatz}
\begin{equation}\label{ctl}
C_{ab}=\lambda W_{a}W_{b},\quad \hat{C}_{ab}=\lambda \left(W_{a}W_{b}-\frac{1}{4}W^{2}g_{ab}\right),\quad C=\lambda W^{2}.
\end{equation}
Here $\lambda$ is a real coupling constant, $W_{a}(x)$ is a smooth field and $W^{2}\equiv W_{a}W^{a}$, as before.

A tedious but straightforward calculation using the above \textit{ansatz} into the characteristic polynomial provided by Eq. (\ref{hxq}) gives 
\begin{equation}
H(x,k)\sim [(h^{-1})^{ab}k_{a}k_{b}]^{2},
\end{equation}
with 
\begin{equation}\label{conteff}
(h^{-1})^{ab}\equiv g^{ab}+\lambda W^{a}W^{b}.
\end{equation}
We notice that this polynomial is not a genuine quartic. In other words, it factorizes as a product of two identical quadratic polynomials. This means that, in this simplified situation, one of the quadratic polynomials must be deleted and the dispersion relation is described by a single symmetric tensor $(h^{-1})^{ab}$, henceforth called the \textit{contravariant effective metric}.

Now, suppose that we manage to find a wave co-vector $k\in T_{x}^{*}M$ that satisfies the dispersion relation
\begin{equation}\label{contradisp}
(h^{-1})^{ab}k_{a}k_{b}=0,
\end{equation}
with $(h^{-1})^{ab}$ provided by Eq. (\ref{conteff}). What can be said about the corresponding polarization co-vector $a\in T_{x}^{*}M$? In order to answer this question, we first notice that the contravariant version of the associated double form can be easily written as
\begin{equation}\label{disprelx}
X^{ambn}=(h^{-1})^{ab}(h^{-1})^{mn}-(h^{-1})^{an}(h^{-1})^{mb}
\end{equation}
i.e. it is a Kulkarni-Nomizu product of the effective metric with itself. Contracting the above expression with the wave co-vector twice and using Eq. (\ref{disprelx}), we get
\begin{equation}
Y^{ab}(k)=X^{ambn}k_{m}k_{n}=-[(h^{-1})^{am}k_{m}][(h^{-1})^{bn}k_{n}],
\end{equation}
which is easily seen to be of rank one. Since the polarization co-vector belongs to the kernel of $Y^{ab}$, the space of nontrivial polarizations must be of dimension two (as in Maxwell's theory) and there follows the ``orthogonality'' condition
\begin{equation}
(h^{-1})^{ab}a_{a}k_{b}=0.
\end{equation}
Thus, both polarization modes travel with precisely the same velocities. Hence, there is no birefringence in this simplified model.

It is easily seen that the inverse of $(h^{-1})^{ab}$ is given by the following \textit{covariant effective metric}
\begin{equation}
h_{ab}=g_{ab}-\lambda \tilde{c}^{2}W_{a}W_{b},
\end{equation}
where
\begin{equation}
\tilde{c}^{2}\equiv \frac{1}{1+\lambda W^{2}}
\end{equation}
plays the role of an ``effective velocity of light'' in a sense which will be described afterwards. With the above convention, a straightforward calculation shows that $(h^{-1})^{ac}h_{cb}=\delta^{a}_{\phantom a b}$ and that the determinants are related by 
\begin{equation}
\mbox{det}(h_{ab})=\tilde{c}^{2}\mbox{det}(g_{ab}).
\end{equation}
Hence, for $h_{ab}$ to have a Lorentzian signature we must impose the algebraic constraint
\begin{equation}
1+\lambda W^{2}>0.
\end{equation}
Roughly speaking, this inequality means that the theory cannot be well defined for arbitrary values of the coupling constant $\lambda$ and the field $W_{a}(x)$. Indeed, if the above relation is somehow violated, there will be a breakdown of hyperbolicity and wave propagation will be ill behaved.
 
In order to better discuss the causal structure of the theory, we need to introduce the concept of a ray. As is well known, for a given $k\in T_{x}^{*}M$ satisfying Eq. (\ref{contradisp}), there will be a ray vector, say $\xi\in T_{x}M$, defined by
\begin{equation}
\xi^{a}\equiv (h^{-1})^{ab} k_{b}
\end{equation}
Clearly, these ray vectors must satisfy the quadratic equation
\begin{equation}\label{dispf}
h_{ab}\xi^{a}\xi^{b}=0, 
\end{equation}
from which we get
\begin{equation}\label{rayeq}
\xi^{2}=\lambda \tilde{c}^{2}(W\cdot\xi)^{2}.
\end{equation}
We notice that for $\lambda\rightarrow 0$, one recovers the usual Minkowskian light cone relation $\xi^{2}=0$. In the following we explore the effective cone structures and the associated constitutive tetrads when $\lambda\neq 0$. Clearly, three distinct possibilities arise due to different characters of $W^{a}$: timelike, null and spacelike. To do so, we consider a spacetime point $x\in M$, a tetrad of vectors such that $g_{ab}|_{x}=\eta_{ab}$ and write the corresponding vectors as $W^{a}=(W^{0},\vec{W})$ and $\xi^{a}=(\xi^{0},\vec{\xi})$. Furthermore, the observer is assumed to be aligned with the first leg of the tetrad i.e. $t^{q}=\delta^{q}_{\phantom a 0}$.  
\subsection{Timelike}
To start with, we consider the simplest possible situation, where the corresponding vector is aligned with the first leg of the tetrad i.e., $W^{a}=X(1,0,0,0)$, for some real number $X$. Plugging the latter into Eq. (\ref{rayeq}), one obtains, after simple manipulations, the ray equation
\begin{equation}
\tilde{c}^{2}(\xi^{0})^{2}-(\xi^{1})^{2}-(\xi^{2})^{2}-(\xi^{3})^{2}=0,
\end{equation}
with
\begin{equation}
\tilde{c}^{2}=\frac{1}{1+\lambda X^2}.
\end{equation}
The above relation shows that the two-dimensional surfaces obtained in $T_{x}M$ by setting $\xi^{0}=const$ are centered spheres, which may
be interpreted as the wave fronts of the electromagnetic field as seen by an observer ``at rest'' with the tetrad. Thus, the excitations travel with the same velocity $\tilde{c}$ in all directions and the intersection between the effective light cone and the ordinary Minkowskian cone is empty (see Figure 1). Also, we identify two distinct regimes associated to the velocity of propagation:
\begin{enumerate}
\item If $\lambda>0$, the effective light velocity is always less than the ordinary velocity of light.
\item If $\lambda<0$, the effective light velocity will be always larger than the ordinary velocity of light.
\end{enumerate}

\begin{figure}
    \centering
    \includegraphics[width=0.35\linewidth]{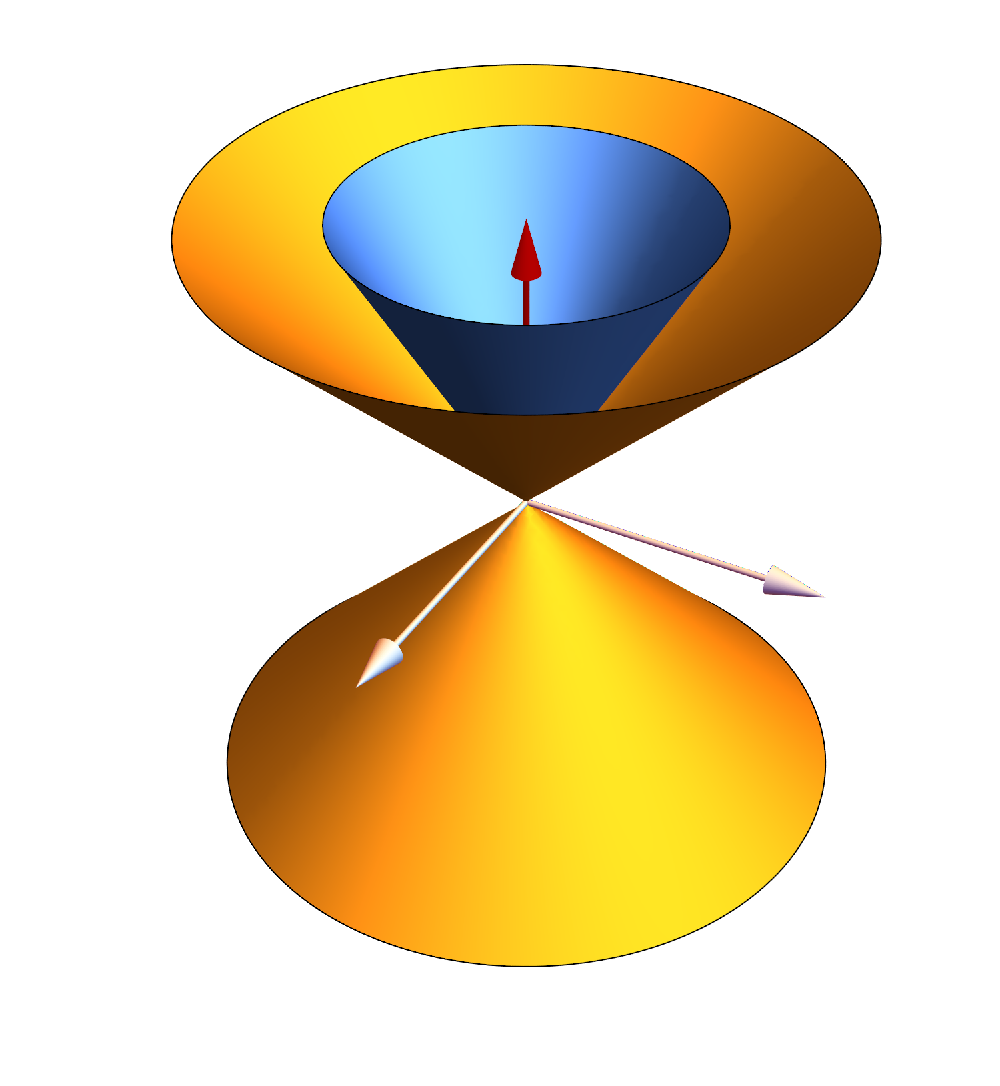}
    \caption{Minkowskian light cone (in yellow), effective light cone for the parameters $\lambda X^{2}=2$ (in blue) and the timelike vector $W^{a}$ (in red).}
    \label{fig:cone1}
\end{figure}

What about the constitutive relations associated to the timelike case? A careful inspection of Eqs. (43-46) give the following constitutive tetrad: 

\begin{equation}
\mathfrak{A}^{i}_{\phantom a j}= (1+\lambda X^{2})\left(\begin{array}{ccc}
1& 0&0 \\
0& 1&0 \\
0 &0 &1 \\
\end{array}\right),\quad\quad\quad \mathfrak{D}^{i}_{\phantom a j}= \left(\begin{array}{ccc}
1& 0&0 \\
0& 1&0 \\
0 &0 &1 \\
\end{array}\right)
\end{equation}
with vanishing magnetoelectric cross terms. This means that the light rays behave as if they were embedded in a linear and isotropic dielectric material with electric permittivity given by $\epsilon=(1+\lambda X^{2})$ and constant magnetic permeability $\mu=1$.

\subsection{Null}

Vectors of a null type are important as they describe possible transitions between timelike regions and spacelike ones. Furthermore, the effective metric corresponding to this case has precisely the form of a Kerr-Schild metric, which is important in the context of black hole physics \cite{Kerr_1963,kerr1965gravitational}. The \textit{ansatz} is provided by $W^{a}=X(1,0,0,1)$ for some real $X$ and the ray equation Eq. (\ref{rayeq}) becomes
\begin{equation}\label{raynull}
\left(1-\lambda X^{2}\right)(\xi^{0})^{2}-(\xi^{1})^{2}-(\xi^{2})^{2}-\left(1+\lambda X^{2}\right)(\xi^{3})^{2}+2\lambda X^{2}\xi^{0}\xi^{3}=0,
\end{equation}
since $W^{2}=0$ in this case. An important feature of this ray equation is that any ray vector of the type $\xi^{a}\sim W^{a}$ automatically satisfies Eq. (\ref{raynull}). This means that, along this direction the effective light cone intersects the ordinary Minkowskian cone (see Figure 2). Clearly, the two-dimensional surfaces obtained in $T_{x}M$ by setting $\xi^{0}=const$ are ellipsoids of revolution which are not centered, but dislocated towards the spatial part of $W^{a}$.

\begin{figure}
    \centering
    \includegraphics[width=0.35\linewidth]{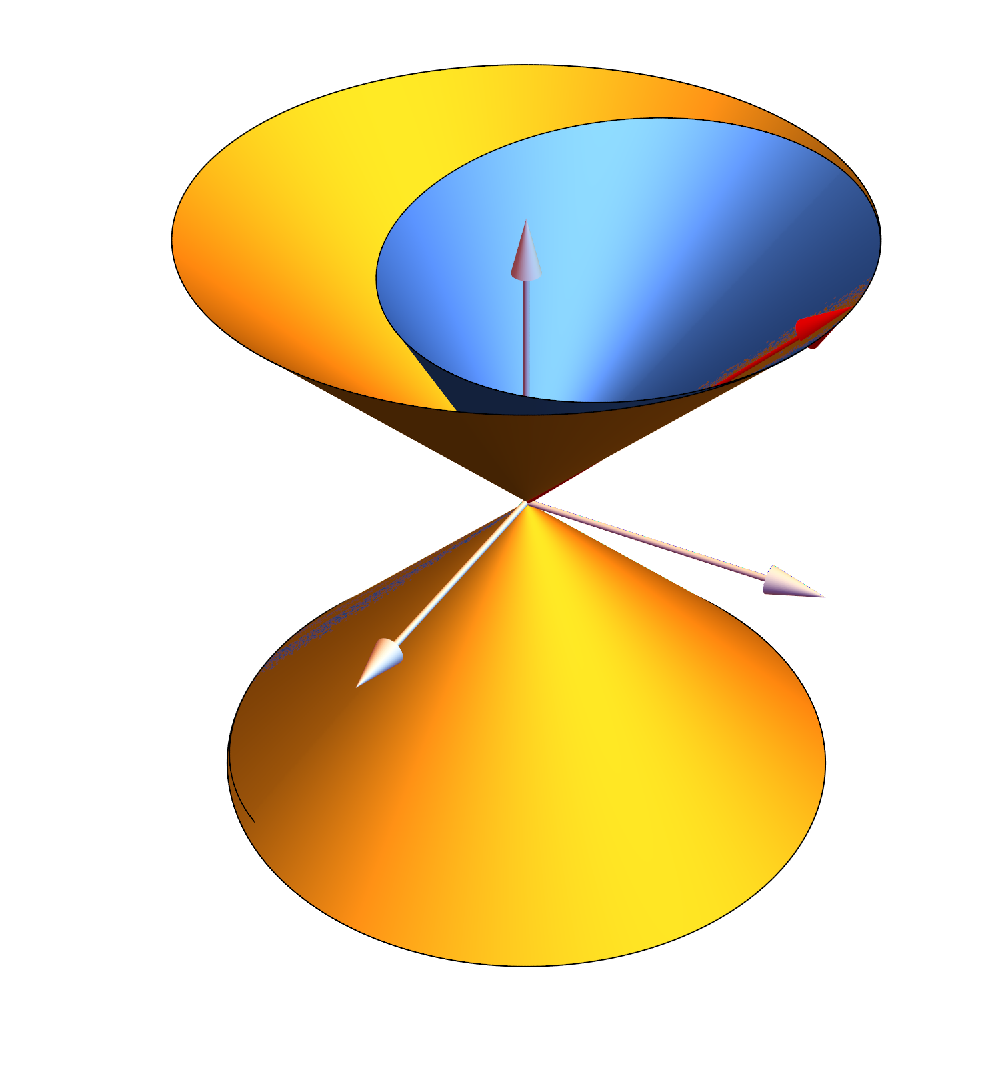}
    \caption{Minkowskian light cone (in yellow), effective light cone  for the parameters $\lambda X^{2}=1/2$ (in blue) and the lightlike vector $W^{a}$ (in red).}
    \label{fig:cone1}
\end{figure}
In this case, Eqs. (43-46) show that the permittivity tensor and the inverse permeability tensor read as
\begin{equation}
\mathfrak{A}^{i}_{\phantom a j}= \left(\begin{array}{ccc}
1+\lambda X^{2}& 0&0 \\
0& 1+\lambda X^{2}&0 \\
0 &0 &1 \\
\end{array}\right),\quad\quad\quad \mathfrak{D}^{i}_{\phantom a j}= \left(\begin{array}{ccc}
1-\lambda X^{2}& 0&0 \\
0& 1-\lambda X^{2}&0 \\
0 &0 &1 \\
\end{array}\right),
\end{equation}
whereas the magnetoelectric cross terms are given by
\begin{equation}
\mathfrak{B}^{i}_{\phantom a j}=\mathfrak{C}^{i}_{\phantom a j}=\left(\begin{array}{ccc}
0& \lambda X^{2}&0 \\
-\lambda X^{2}& 0&0 \\
0 &0 & 0\\
\end{array}\right).
\end{equation}
One then notices that, although $\mathfrak{A}^{i}_{\phantom a j}$ and $\mathfrak{D}^{i}_{\phantom a j}$ are simultaneously in diagonal form, their eigenvalues are not equal. This feature causes anisotropies in the propagation that are reinforced by the magnetoelectric cross terms. The latter somehow drags the effective light cone in the direction of $W^{a}$, very much in the same way as sound waves would behave in transonic drifting water.

\subsection{Spacelike}

By choosing an orthonormal frame such that the third axis is aligned with the vector, we write $W^{a}=X(0,0,0,1)$ and there follows $W^2=-X^{2}$. In this case, Eq. (\ref{dispf}) becomes
\begin{equation}
(\xi^{0})^{2}-(\xi^{1})^{2}-(\xi^{2})^{2}-\tilde{c}^{2}(\xi^{3})^{2}=0
\end{equation}
with 
\begin{equation}
\tilde{c}^{2}\equiv\frac{1}{1-\lambda X^{2}}
\end{equation}
Thus, along all directions orthogonal to the spacelike vector $W^{a}$ the field excitations propagate with the usual velocity of light $``c"$, whereas in all other directions the velocity is modified.

The constitutive tetrad is given by
\begin{figure}
    \centering
    \includegraphics[width=0.35\linewidth]{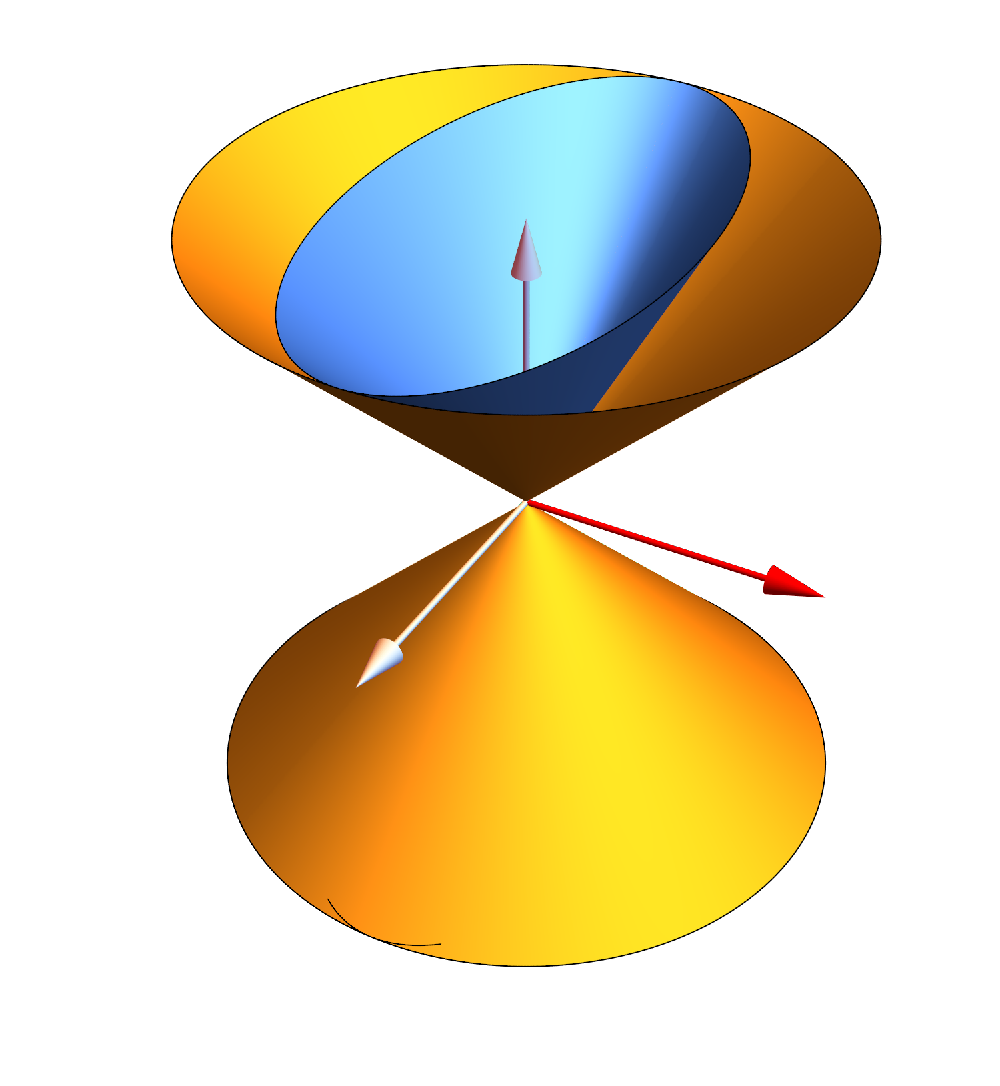}
    \caption{Minkowskian light cone (in yellow), effective light cone  for the parameters $\lambda X^{2}=1/2$ (in blue) and the spacelike vector $W^{a}$ (in red).}
    \label{fig:cone1}
\end{figure}

\begin{equation}
\mathfrak{A}^{i}_{\phantom a j}= \left(\begin{array}{ccc}
1& 0&0 \\
0& 1&0 \\
0 &0 & 1-\lambda X^{2}  \\
\end{array}\right),\quad\quad\quad \mathfrak{D}^{i}_{\phantom a j}= \left(\begin{array}{ccc}
1-\lambda X^{2}& 0&0 \\
0& 1-\lambda X^{2}&0 \\
0 &0 &1 \\
\end{array}\right),
\end{equation}
with vanishing magneto-electric cross terms. Again, $\mathfrak{A}^{i}_{\phantom a j}$ and $\mathfrak{D}^{i}_{\phantom a j}$ are simultaneously in diagonal form, but their eigenvalues are not equal. In its turn, this fact implies in anisotropic propagation: the two-dimensional surfaces obtained in $T_{x}M$ by setting $\xi^{0}=const$ are centered ellipsoids of revolution about $W^{a}$.

\section{Conclusion}

 In this paper, we have performed a systematic treatment of light propagation in the context of Lorentz-breaking electrodynamics. The study is important mainly because other extensions of Maxwell's theory such as nonlinear electrodynamics, non-minimal couplings with the gravitational field and area metric theories also predict a modification of the ordinary dispersion relation. Therefore, in order to test Lorentz-breaking electrodynamics in extreme astrophysical and cosmological settings it is mandatory from the experimental point of view that we manage to separate and categorize all possible effects engendered by the different theories. Since the Standard Model Extension program is too general for this purpose, we have concentrated our analysis into a specific class of the CPT-even sector. 

 Relying on a fully covariant approach, we have explicitly derived the dispersion relation associated to the class of models. We have shown that the characteristic polynomial is necessarily of fourth order in the wave covector and contains up to third order terms in the Lorentz-breaking tensor field $\hat{C}_{ab}(x)$. In practice, this means that the photons of the theory do not propagate along the ordinary Minkowskian light cone, but rather on a modified effective cone which may contain self-intersections and singular regions. Furthermore, using the Bel-Matte decomposition of double forms, we have obtained the corresponding constitutive tetrad associated to the class of models. This is important because the constitutive tetrad permits us to map the Lorentz breaking scenario into light propagating in anisotropic material media, which may be easily tested in a laboratory framework.

 In order to check the geometry of the effective cones in more details, we then considered a simplified model where $\hat{C}_{ab}$ has a trivial dyadic structure induced by a vector field $W^{a}$. In this case, we have managed to show that the fourth order disperion relation factorizes into a product of identical quadratic ones and bi-refringence is absent. Roughly speaking, this means that the propagation in this case does not depend on the polarization of the wave. Finally, we have studied how the cones are modified according to the character of the underlying vector: timelike, lighlike and spacelike. We showed that each case presents its own subtleties and may be used to test the model: the first case corresponds to an isotropic dielectric medium whereas the other two correspond to different types of anisotropic material media.   
%
%
%
\

\bibliographystyle{unsrt}
\bibliography{biblio.bib}

\begin{thebibliography}{10}

\bibitem{kost1}
V.~Alan Kosteleck\'y and Stuart Samuel.
\newblock Spontaneous breaking of lorentz symmetry in string theory.
\newblock {\em Phys. Rev. D}, 39:683--685, Jan 1989.

\bibitem{colladay1997cpt}
Don Colladay and V~Alan Kosteleck{\`y}.
\newblock Cpt violation and the standard model.
\newblock {\em Physical Review D}, 55(11):6760, 1997.

\bibitem{colladay1998lorentz}
Don Colladay and V~Alan Kosteleck{\`y}.
\newblock Lorentz-violating extension of the standard model.
\newblock {\em Physical Review D}, 58(11):116002, 1998.

\bibitem{kostelecky2004gravity}
V~Alan Kosteleck{\`y}.
\newblock Gravity, lorentz violation, and the standard model.
\newblock {\em Physical Review D}, 69(10):105009, 2004.

\bibitem{sorokin2022introductory}
Dmitri~P Sorokin.
\newblock Introductory notes on non-linear electrodynamics and its applications.
\newblock {\em Fortschritte der Physik}, 70(7-8):2200092, 2022.

\bibitem{balakin2005non}
Alexander~B Balakin and Jos{\'e}~PS Lemos.
\newblock Non-minimal coupling for the gravitational and electromagnetic fields: a general system of equations.
\newblock {\em Classical and Quantum Gravity}, 22(9):1867, 2005.

\bibitem{punzi2009propagation}
Raffaele Punzi, Frederic~P Schuller, and Mattias~NR Wohlfarth.
\newblock Propagation of light in area metric backgrounds.
\newblock {\em Classical and Quantum Gravity}, 26(3):035024, 2009.

\bibitem{perlick2011hyperbolicity}
Volker Perlick.
\newblock On the hyperbolicity of maxwell's equations with a local constitutive law.
\newblock {\em Journal of mathematical physics}, 52(4), 2011.

\bibitem{Raetzel:2010je}
Dennis Raetzel, Sergio Rivera, and Frederic~P. Schuller.
\newblock {Geometry of physical dispersion relations}.
\newblock {\em Phys. Rev. D}, 83:044047, 2011.

\bibitem{de2000light}
VA~De~Lorenci, Renato Klippert, M~Novello, and JM~Salim.
\newblock Light propagation in non-linear electrodynamics.
\newblock {\em Physics Letters B}, 482(1-3):134--140, 2000.

\bibitem{abalos2015nonlinear}
Fernando Abalos, Federico Carrasco, {\'E}rico Goulart, and Oscar Reula.
\newblock Nonlinear electrodynamics as a symmetric hyperbolic system.
\newblock {\em Physical Review D}, 92(8):084024, 2015.

\bibitem{de2009classification}
{\'E}rico~Goulart de~Oliveira~Costa and Santiago Esteban~Perez Bergliaffa.
\newblock A classification of the effective metric in nonlinear electrodynamics.
\newblock {\em Classical and Quantum Gravity}, 26(13):135015, 2009.

\bibitem{Goulart:2021uzr}
{\'E}rico Goulart and Santiago Esteban~Perez Bergliaffa.
\newblock {Nonlinear electrodynamics nonminimally coupled to gravity: Symmetric-hyperbolicity and causal structure}.
\newblock {\em Phys. Rev. D}, 105(2):024021, 2022.

\bibitem{Carroll}
S.~Carroll and H.~Tam.
\newblock {\em Physical Review D}, 78:044047, 2008.

\bibitem{Petrov}
M.~Gomes, J.~R. Nascimento, A.~Yu. Petrov, and A.~J. da~Silva.
\newblock {\em Physical Review D}, 81:045018, 2010.

\bibitem{Scarp}
G.~Gazzola, H.~G. Fargnoli, A.~P.~Baêta Scarpelli, Marcos Sampaio, and M.~C. Nemes.
\newblock {\em Journal of Physics G}, 39:035002, 2012.

\bibitem{Scarp2}
A.~P.~Baêta Scarpelli.
\newblock {\em Journal of Physics G}, 39:125001, 2012.

\bibitem{Petrov-Scarp}
A.~P.~Baêta Scarpelli, T.~Mariz, J.~R. Nascimento, and A.~Yu. Petrov.
\newblock {\em European Physical Journal C}, 73:2526, 2013.

\bibitem{Kost5}
V.~A. Kostelecký and N.~Russell, 2008.
\newblock arXiv:0801.0287 [hep-ph].

\bibitem{Kost-CPT}
V.~Alan Kostelecký, 2022.
\newblock arXiv:2210.09824 [hep-ph].

\bibitem{Adda}
A.~Addazi et~al.
\newblock {\em Progress in Particle and Nuclear Physics}, 124:103948, 2022.

\bibitem{Kost2}
V.~A. Kostelecký and M.~Mewes.
\newblock {\em Physics Letters B}, 757:510, 2016.

\bibitem{Kost3}
V.~A. Kostelecký, A.~C. Melissinos, and M.~Mewes.
\newblock {\em Physics Letters B}, 761:1, 2016.

\bibitem{Schr}
M.~Schreck.
\newblock {\em Classical and Quantum Gravity}, 34:135009, 2017.

\bibitem{Kost4}
V.~A. Kostelecký and J.~D. Tasson.
\newblock {\em Physics Letters B}, 749:551, 2015.

\bibitem{Alts}
B.~Altschul.
\newblock {\em Physical Review D}, 78:085018, 2008.

\bibitem{Bert}
O.~Bertolami and C.~S. Carvalho.
\newblock {\em Physical Review D}, 61:103002, 2000.

\bibitem{Dopli}
S.~Doplicher, K.~Fredenhagen, and J.~E. Roberts.
\newblock {\em Communications in Mathematical Physics}, 172:187, 1995.

\bibitem{Boj}
M.~Bojowald, H.~A. Morales-Tecotl, and H.~Sahlmann.
\newblock {\em Physical Review D}, 71:084012, 2005.

\bibitem{Amel}
G.~Amelino-Camelia.
\newblock {\em Living Reviews in Relativity}, 16:5, 2013.

\bibitem{goulart2023remarks}
E.~Goulart and J.~E. Ottoni.
\newblock Remarks on the algebraic structure of (2,2) double forms.
\newblock {\em International Journal of Geometric Methods in Modern Physics}, 22(08):2550053, 2025.

\bibitem{hehl2012foundations}
Friedrich~W Hehl and Yuri~N Obukhov.
\newblock {\em Foundations of classical electrodynamics: Charge, flux, and metric}, volume~33.
\newblock Springer Science \& Business Media, 2012.

\bibitem{itin2009light}
Yakov Itin.
\newblock On light propagation in premetric electrodynamics: the covariant dispersion relation.
\newblock {\em Journal of Physics A: Mathematical and Theoretical}, 42(47):475402, 2009.

\bibitem{lindell2005electromagnetic}
Ismo~Veikko Lindell.
\newblock Electromagnetic wave equation in differential-form representation.
\newblock {\em Progress In Electromagnetics Research}, 54:321--333, 2005.

\bibitem{favaro2012recent}
Alberto Favaro.
\newblock {\em Recent advances in classical electromagnetic theory}.
\newblock PhD thesis, Imperial College London, 2012.

\bibitem{Kerr_1963}
Roy~P. Kerr.
\newblock Gravitational field of a spinning mass as an example of algebraically special metrics.
\newblock {\em Phys. Rev. Lett.}, 11:237--238, Sep 1963.

\bibitem{kerr1965gravitational}
Roy~Patrick Kerr.
\newblock Gravitational collapse and rotation.
\newblock {\em Quasi-Stellar Sources and Gravitational Collapse}, page~99, 1965.

\end{thebibliography}

\end{document}